\documentclass{aa} 

\usepackage{graphicx}
\usepackage{txfonts}
\usepackage[citecolor=blue, linkcolor=blue, urlcolor = black, colorlinks = true]{hyperref}

\usepackage{graphicx}	
\usepackage{amsmath}	
\usepackage{amssymb}	
\usepackage{xspace}

\usepackage{siunitx}
\usepackage{comment}
\usepackage{textcomp}
\usepackage{bm}
\usepackage{lscape}
\usepackage{xcolor}
\usepackage{natbib}
\usepackage{subcaption}

\newcommand{\Msun}{\,M$_{\odot}$\xspace}

\newcommand{\fov}{$f_{\rm CBM}$\xspace}

\newcommand{\percent}{~per~cent\xspace}

\newcommand{\edit}[1]{#1} 

\newcommand{\mesa}{\texttt{MESA}\xspace}

\newcommand{\varghese}{V23\xspace}

\usepackage{hyperref}
\begin{document}

   \title{Calibrating chemical mixing induced by internal gravity waves based on hydrodynamical simulations }
   \subtitle{The chemical evolution of OB-type stars}

   \author{J.~S.~G. Mombarg\inst{1,2}
          \and
          A. Varghese\inst{2,3}
          \and 
          R.~P. Ratnasingam\inst{3}
          }

   \institute{IRAP, Universit\'e de Toulouse, CNRS, UPS, CNES, 14 avenue \'Edouard Belin, F-31400 Toulouse, France\\
              \email{joey.mombarg@cea.fr}
              \and 
              {Universit\'e Paris-Saclay, Universit\'e de Paris, Sorbonne Paris Cit\'e, CEA, CNRS, AIM, 91191 Gif-sur-Yvette, France} 
              \and 
              {School of Mathematics, Statistics and Physics, Newcastle University, Newcastle upon Tyne, NE1 7RU, UK}
        }

   \date{Received November 11 2024; accepted February 18 2025}
\titlerunning{Stellar evolution with IGW mixing}
\authorrunning{Mombarg, Varghese \& Ratnasingam}
 
  \abstract
   {Internal gravity waves (IGWs) have been shown to contribute to the transport of chemical elements in stars with a convective core and radiative envelope. Recent two-dimensional hydrodynamical simulations of convection in intermediate-mass stars have provided estimates of the chemical mixing efficiency of such waves. The chemical diffusion coefficient from IGW mixing is described by a constant $A$ times the squared wave velocity. The value of $A$, however, remains unconstrained by such simulations. 
   }
   {This work aims at investigating what values $A$ can take in order to reproduce the observed nitrogen surface abundances of the most nitrogen-enriched massive stars. Furthermore, we discuss the prevalence of IGW mixing compared to rotational mixing.   }
   {We provide an implementation of these (time-dependent) mixing profiles predicted from hydrodynamical simulations in the one-dimensional stellar evolution code \mesa. We compute evolution tracks for stars between 3 and 30\Msun with this new implementation for IGW mixing and study the evolution for the surface abundances of isotopes involved in the CNO cycle, particularly the nitrogen-14 isotope.}
   {We show that this one-dimensional framework predicting the chemical diffusion coefficient from IGW mixing yields consistent \edit{morphologies of the mixing profile} in comparison with hydrodynamical simulations. We find that the value of $A$ must increase with mass in order to reproduce the most nitrogen-enriched stars. \edit{Assuming these calibrated values for $A$, mixing by IGWs is a potential mechanism to reproduce well-mixed stars without needing rapid rotation.}  }
   {We have provided observational limits on the efficiency of IGW mixing for future theoretical studies. Furthermore, future asteroseismic modelling efforts taking IGW mixing into account will be able to place additional constraints on the convective core mass, as our models predict that the convective core should be significantly more massive if IGW mixing is indeed efficient. }


   \maketitle
%
\section{Introduction}
One of the most influential, yet unconstrained, physical ingredients in the state-of-the-art stellar structure and evolution models is the transport of chemical species throughout the star, also referred to as (chemical) mixing (see \citet{Salaris2017} for a summary). For intermediate-mass and massive stars, which have a convective core and a radiative envelope, the time scale(s) of the mixing greatly impacts a star's evolutionary pathway and the chemical enrichment of the Universe \citep{Langer2012}. Presently, the need for chemical mixing inside the radiative zones has been well established by observational characterisation of the chemical compositions of stellar surfaces \citep[e.g.][]{Hunter2008,Hunter2009, Brott2011a, Brott2011b,Maeder2014,Martins2017,Gebruers2021}, as well as asteroseismic modelling of stellar interiors (e.g. \citealt{Pedersen2021, Mombarg2022, Burssens2023}, and \citealt{Bowman2020-review} for a review). The prime mechanisms by which material can be transported in the radiative zones are thought to be instabilities that drive turbulence as a result of differential rotation \citep{Zahn1992, Heger2000, Maeder2003}, and transport induced by internal gravity waves (IGWs) excited at interfaces of convective and radiative regions \citep{Montalban1994, Montalban1996, RogersMcElwaine2017, Herwig2023}.

The implementation of both these mechanisms in one-dimensional stellar evolution codes is not a trivial matter, and while stellar evolution models that include rotationally induced mixing have been successful in reproducing previously ill-understood observed surface compositions, these models do not provide a full explanation for the observed scatter of the full distribution of the surface nitrogen abundances and the projected surface rotational velocity or the near-core rotation frequency \citep{Brott2011b, aerts2014}. Stellar evolution models that include the effects of fossil magnetic field are able to reproduce stars with a slowly-rotating nitrogen-enriched surfaces, as a result of magnetic braking \citep{Meynet2011, Keszthelyi2020}. This might imply a higher incidence rate of fossil magnetic fields in lower-metallicity environments, compared to the observed rate around 10\percent for Galactic OBA-type dwarfs \citep[e.g.][]{Grunhut2017}. Furthermore, a qualitative comparison between models and spectroscopic observations comes with intricacies, such as the uncertainty on the stellar mass determination.

Hydrodynamical simulations of the 
stellar interiors in two and three dimensions have provided us with better insight into the wave excitation, either by bulk excitation by Reynolds stresses, or by convective plumes, and the wave propagation in the stably stratified region \citep{belkam_2009, pincon2016}. The predicted morphology of IGW frequency spectra by two-dimensional (2D) simulations \citep{Horst2020, Ratnasingam2020} and 3D simulations \citep{Edelmann2019, Vanon2023, Thompson2024} have been shown to be in agreement with observational signatures \citep{Bowman2019-aa, Bowman2019-nature, Bowman2020, BowmanDorn2022}. Furthermore, \cite{Rogers2015} showed that predicted rotation profiles from 2D simulations are in line with asteroseismic measurements of the near-core rotation frequency, although the constraining power of these measurements was limited. Introducing tracer particles in such simulations of stars with a convective core and radiative envelope have revealed that the mixing\footnote{The transport of angular momentum by IGWs can, however, not be treated as diffusive, as IGWs induce shear flows.} can be treated as a diffusive process \citep{RogersMcElwaine2017}. This implies that the efficiency of the mixing driven by IGWs can be characterised by a chemical diffusion coefficient. The influence of IGWs excited by bulk excitation have been accounted for in the 1D stellar structure and evolution code \texttt{STAREVOL} \citep{Charbonnel2013}. The influence of IGWs on the chemical mixing in this case is however only accounted for indirectly by the feedback of waves on the rotation profile, and thereby the rotationally-induced processes. Yet, the indirect contribution of IGWs to the mixing by angular momentum transport has been shown to yield consistent predictions of Li abundances in solar-type stars \citep{Charbonnel2005,Talon2005}. \edit{This does, however, not imply that the theory of wave mixing is complete, as we also expect the IGWs to have a direct contribution to the chemical mixing \citep{Varghese2023}. In this paper, we aim to study the direct contribution of IGWs, focussing on stars with a convective core and radiative envelope.} 

So far, studies of massive stars that do include models with IGW mixing informed by hydrodynamical simulations typically scale the diffusion coefficient with the local density, where the dependence as somewhere between $\rho^{-1/2}$ and $\rho^{-1}$ \citep{RogersMcElwaine2017}. One such study, \cite{Varghese2023}, presents the
predicted chemical mixing induced by IGWs by running tracer particle simulations using the velocity field data
from 2D hydrodynamical simulations as a post-process step.

For \edit{3, 7 and 20\Msun}, three points along the main sequence were considered; close to zero-age main sequence (ZAMS), mid-age main sequence (MAMS), and close to terminal-age main sequence (TAMS). An important result from this study is the significant change of the IGW mixing profile throughout the main sequence. In younger stars, they found that the diffusion profile follows an increase trend
towards the surface whilst in older stars, the mixing profile shows an initial increase followed by a decreasing
trend towards the surface. The differences here can be largely attributed to the interaction between
the waves and the background stratification. 

In this paper, we present a 1D prescription for the chemical mixing generated by IGWs that accounts for the time evolution seen in these 2D simulations, and study the effect on the chemical evolution. \edit{We observationally calibrate a parameter that cannot be constrained from hydrodynamical simulations.}   
The paper is structured as follows. We describe the numerical approach of implementing IGW mixing into a 1D stellar structure and evolution code in Sect.~\ref{sec:IGW}. We then discuss the predicted evolution of the surface abundances and convective core mass in Sect.~\ref{sec:results} and Sect.~\ref{sec:core_masses}, respectively. In Sect.~\ref{sec:add_mix}, we discuss other sources of chemical mixing, and we conclude in Sect.~\ref{sec:conclusions}.

\section{One-dimensional implementation of wave mixing} \label{sec:IGW}
In this paper, we provide a 1D implementation to include the direct contribution of IGWs to the transport of chemical species, relying on the results of \citet[][hereafter \varghese]{Varghese2023}. \edit{The implementation presented in this section is based on their results for the 3\Msun and 7\Msun models, but not the 20\Msun models. The exclusion of the 20\Msun model is done here because it is less representative of realistic stars of similar or higher masses compared to the other two masses, due to several factors. Firstly, because the internal structure of the 20\Msun model is significantly different compared to the other two masses, the model did not achieve full convergence within the time frame of the simulation. Secondly, the model was cut off at 80\percent of the total stellar radius for numerical stability, rather than at 90\percent. Lastly, at such high masses and above, we expect stars to have significantly larger outer convection zones and shear flows near the surface. 
Thus, we omit the discussion of the 20\Msun model in this work. One final note is that we study the isolated effect of IGW mixing, and therefore do not account for interactions between the waves and the rotation. }
 
 \subsection{Numerical approach} \label{sec:num_approach}
 The 2D simulations informing the work in this paper
are detailed in \varghese and \cite{rathish_2023}. The techniques themselves are elaborated in \cite{rogers_2013}.
Therefore, here, we only provide a summary of these simulations relevant to this work. The simulations solve the
anelastic Navier-Stokes equations within the geometry of an equatorial slice.
The simulation domains extend up to 0.9\,R$_{\star}$ (stellar radii) and in terms of age, three points along the main sequence are
considered; close \edit{to ZAMS, MAMS, and close to TAMS.} These simulations capture the global internal structure evolution of intermediate-mass stars within convective timescales as the convective processes within the convection zone self-consistently generate gravity
waves that propagate to the top of the radiation zone. The reference state variables used for these simulations are available on Zenodo\footnote{\url{https://zenodo.org/record/2596370\#.Yn5quDnMJUR}}.

 Following the above 2D formalism, we scale the chemical diffusion coefficient in the radiative envelope with the squared wave amplitude as per
\begin{equation}\label{eqn:param_D}
    D_{\rm IGW}(r) = A \sum_m^{m_{\rm max}} \sum_{\omega_i} v^2_{\rm wave}(\omega_i, m, r),
\end{equation}
where $A$ is a constant with the units in $s$ as expected from the Eq.~(\ref{eqn:param_D}). \edit{The value of $A$ cannot be constrained from hydrodynamical simulations alone and in this work, we aim to calibrate its value to observed surface abundances of massive stars.} 
Here, we only consider 2D Fourier basis wavenumber $m_{\rm max} = 1$, as this choice most accurately reproduces the decrease of the diffusion coefficient towards the surface that is observed in the near-TAMS mixing profiles of \varghese. The frequency that dominates the total diffusion coefficient is dependent on the stellar mass and age and currently there exists no prescription to predict its value accurately. Therefore, we take a sum over \edit{integer frequencies within the range} $\omega_i/(2 \pi) \in [4, 5, \dots, 14\,\mu{\rm Hz}]$, where the range corresponds the most dominant frequencies found by \varghese. This means we study the maximised effect.

The wave amplitude is given by \citep{Ratnasingam2019},
\begin{equation} \label{eq:vwave}
    v_{\rm wave}(\omega, m, r) = \bar{v}_{\rm conv} \left( \frac{\rho}{\rho_0}\right)^{-\frac{1}{2}} \left( \frac{r}{r_0}\right)^{-1} \left( \frac{N^2 - \omega^2}{N_0^2 - \omega^2}\right)^{-\frac{1}{4}} e^{-\frac{\mathcal{T}}{2}},
\end{equation}
where $\bar{v}_{\rm conv}$ is the radially averaged convective velocity according to mixing length theory (MLT), $\omega$ the frequency of the IGW. 
\edit{The MLT convective velocity following \citet{Cox1968} is determined by,
\begin{equation}
    v_{\rm conv} = \alpha_{\rm MLT} \sqrt{\frac{QP}{8 \rho}} \frac{\Gamma}{A_{\rm cond}},
\end{equation}
where $\alpha_{\rm MLT}$ is the MLT parameter, $Q = \chi_T/\chi_\rho$, $P$ is the local pressure, $\Gamma$ the convective efficiency, and $A_{\rm cond}$ is the ratio of convective to radiative conductivity.

In reality, only a fraction of the convective plume is transmitted into the IGW, and therefore the velocity that we use here is the upper limit. The computation of the transmitted velocity is, however, numerically ill-defined (see Appendix~\ref{ap:uv0}), and therefore we absorb this uncertainty into the parameter $A$. 
}

As this work is driven mainly by the comparison to 2D hydrodynamical simulations on the equatorial plane, the radius dependency in Eq.~(\ref{eq:vwave}) goes as $r^{-1}$. This would be $r^{-3/2}$ in 3D.
The damping rate $\mathcal{T}(\omega, m, r)$ is computed according to \citet{Kumar1999}:

\begin{equation}
    \mathcal{T}(\omega, m, r) = \frac{16}{3}\sigma_{\rm SB}\int_{r_0}^r \frac{ T^3}{\kappa_{\rm R} \rho^2 C_P}\left( 1 - \frac{\omega^2}{N^2} \right) \left( \frac{k_h^{3} N^3}{ \omega^4}\right) {\rm d}r,    
\end{equation}
where $\sigma_{\rm SB}$ is the Stefan-Boltzmann constant, $T$ is the local temperature, $\kappa_{\rm R}$ is the local Rosseland mean opacity, $C_P$ is the specific heat at constant pressure, $k_h=m/r$ is the horizontal wavenumber.

\begin{figure}
    \centering
    \includegraphics[width = 0.95\columnwidth]{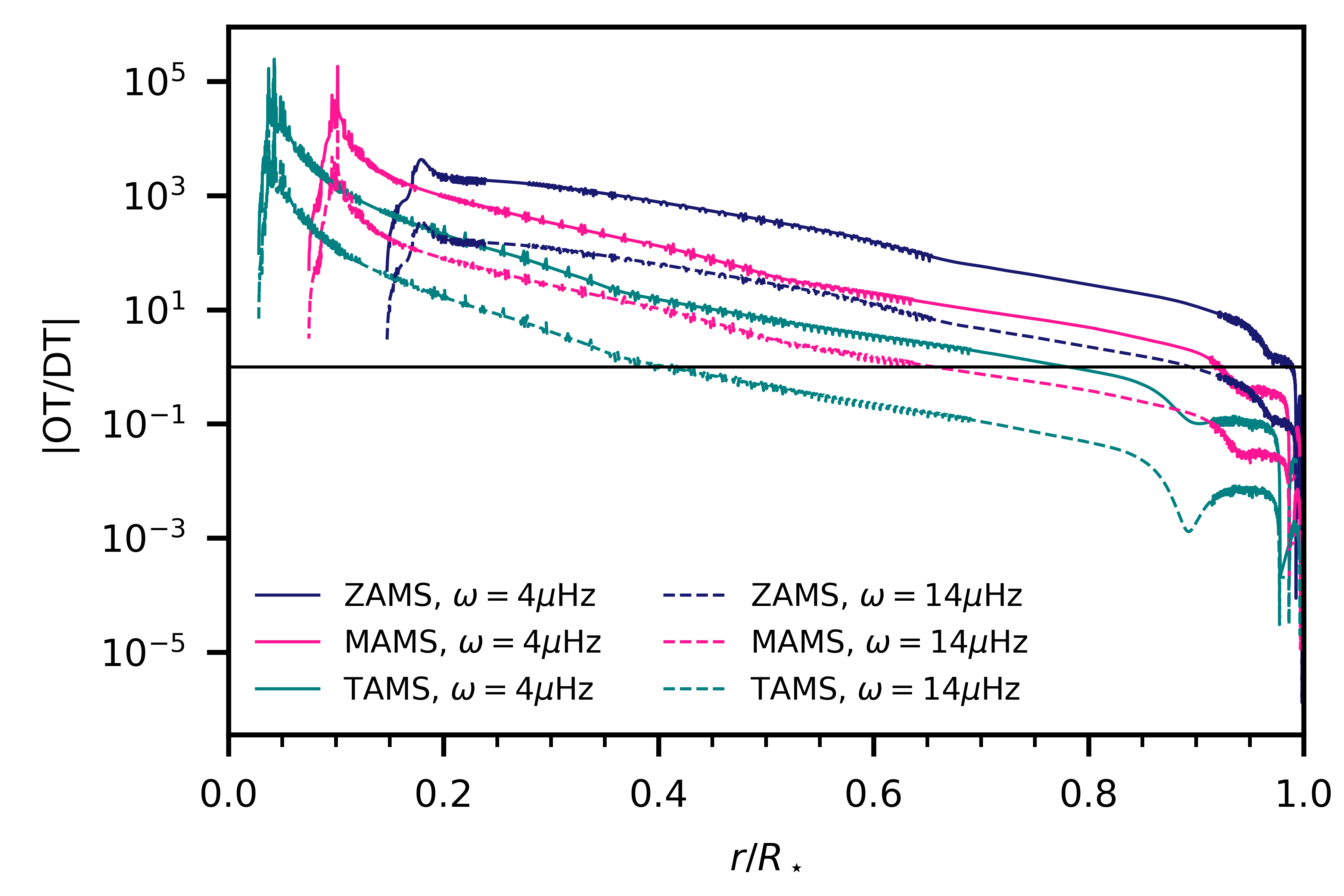}
    \caption{Absolute ratio of the oscillatory term (OT) and density term (DT) in Eq.~(\ref{eq:turn_point}) as a function of radial coordinate. The IGWs experience additional damping when this ratio is below one (indicated by the solid horizontal line). The ratio is shown for two different IGW frequencies from a 3\Msun model with $Z = 0.02$.  }
    \label{fig:OTDT}
\end{figure}

While \varghese used a constant thermal diffusivity in their simulations, we compute the thermal diffusivity from a stellar structure and evolution model (not constant throughout the star). \edit{As mentioned in \varghese, the dominant factors influencing the trend of the radial diffusion profile are the Brunt-V\"ais\"al\"a frequency, and the location of the turning point beyond which the waves lose their wave-like behaviour as seen in older stars. Hence, we expect the thermal diffusivity to not have an influence on the morphology of the profile}. The quantities indicated by a subscript `0' are evaluated at a reference radius $r_0 = r_{\rm cc} + 0.02 h_P(r_{\rm cc})$ taken as the wave launch point, where $r_{\rm cc}$ is the radius of the convective core. The smaller value of 0.02 is used, compared to the values used by \varghese (1.3 times the overshoot depth), to ensure that the wave launching point is well within the core-boundary mixing zone, even for the lowest values of \fov. \varghese found that this analytical paradigm of wave mixing (Eq.~(\ref{eqn:param_D})) does not work well for the stellar models near the end of the main sequence, as IGWs experience additional damping beyond its turning point \citep{Ratnasingam2020}, defined as the radius at which, 

\begin{equation} \label{eq:turn_point}
\frac{1}{2}\left( \frac{\partial h_\rho}{\partial r} - \frac{h_\rho^2}{2} + \frac{h_\rho
}{r}\right) = \left( \frac{N^2}{\omega^2} - 1 \right) \frac{m^2}{r^2},
\end{equation}

\noindent where the LHS corresponds to the density term (DT) and the RHS corresponds to the oscillatory term (OT). When the ratio |OT/DT| is less than one, waves are rapidly attenuated as they become evanescent, which occurs to a fraction of them as they propagate towards the stellar surface.
In this criterion, $h_\rho = -\frac{\partial \ln \rho}{\partial r}$ (i.e. the negative inverse density scale height). 
Figure~\ref{fig:OTDT} shows the ratio of the OT and the DT for a 3\Msun at ZAMS, MAMS and TAMS at two different frequencies (4 and 14\,$\mu$Hz).
In the 2D simulations, the radius at which the diffusion coefficient turn-off
starts to occur is mainly determined by the contribution from a spectrum of waves and their turning points (see
dashed-dotted lines in Fig.~\ref{fig:D_IGW_profiles}). In our 1D implementation, we only account for the dominant frequencies, which
are mainly the lowest frequencies, from the overall spectrum, which shift the diffusion coefficient turn-off radius
to lower values.

We assume that the wave velocity decreases exponentially as a function of $r$ beyond the turning point $r_{\rm t}$, then
\begin{equation}
    D_{\rm IGW}(r>r_{\rm t}) = Av^2_{\rm wave}(\omega, m, r)\exp \left( \frac{-(r - r_{\rm t})}{f R_\star}\right).
\end{equation}
The length scale of the damping is set by a free parameter $f$, for which we find $f = 0.025$ results in a similar decrease of the chemical diffusion coefficient beyond the turning point compared to the results from \varghese. 
 The value of the constant $A$ was found to be around $\sim 1\,{\rm s}$ found by these authors. However, contrary to simulations presented in \varghese and \cite{RogersMcElwaine2017}, in this work we do not use a constant thermal diffusivity to compute the damping rate, and therefore need to change the value of $A$ to obtain similar chemical diffusion coefficients.

 In addition, the radially-averaged convective velocities in the core of the \mesa model are about an order of magnitude smaller than those from the 2D hydrodynamical simulations described at the beginning of this section. The average convective velocity predicted by MLT in the convective core increases as the core shrinks, while the simulations for \varghese show the opposite. The simulations of \varghese most likely do not cover enough convective turnover times for the global amplitude of the mixing profile to have fully converged. This is because such hydrodynamical simulations require long computation times. Therefore, it is only useful for us to reproduce the radial dependence of the mixing profiles as a function of age, thus leaving the constant $A$ an uncalibrated parameter at this point. Finally, we smooth the resulting mixing profile to better reproduce the behaviour of $D_{\rm IGW}$ around the wave turning points and do not change the local diffusion coefficient in convection zones. 

 \edit{One limitation of this current implementation, as previously mentioned, is that it does not account for IGWs that are excited by the possible presence of an outer convective layer and their interaction with outward propagating waves. Future hydrodynamical simulations studying these interactions will provide a more complete picture.  }

\subsection{Stellar structure and evolution models}
In this paper, we aim to study the chemical evolution of stars that experience mixing via IGWs excited at the convective core boundary and propagate through the radiative envelope. Relying on the predicted time evolution of the IGW mixing profile by 2D hydrodynamic simulations by \cite{Varghese2023}, we present a 1D implementation in the stellar structure and evolution code \mesa \texttt{r24.03.1} \citep{Paxton2011, Paxton2013, Paxton2015, Paxton2018, Paxton2019, Jermyn2023}. The initial helium mass fraction in our \mesa models are scaled with the initial metallicity as per $Y_{\rm ini} = Y_{\rm P} + 1.226\,Z_{\rm ini}$, where $Y_{\rm P} = 0.244$ \citep{Verma2019}. We investigate three different metallicity environments corresponding to the Small Magellanic Cloud (SMC), the Large Magellanic Cloud (LMC), and the Milky Way (MW). For the SMC and LMC models, we follow a similar approach as \cite{Keszthelyi2020}, where we initialise the models with custom fractions for the relative metal fractions corresponding to the SMC and LMC baselines given by \cite{Dopita2019}. For the MW models, the initial relative metal mass-fractions correspond to those of the Sun as measured by \cite{Asplund2009} with modifications to several elements following \cite{Nieva2012}. 

The Rosseland mean opacity is computed from the data provided by the OP project \citep{Seaton2005}. In addition to the mixing in the radiative envelope, convective boundary mixing (CBM) is included, where the length scale of the overshoot zone, a fraction \fov of the local pressure scale height, is left as a free parameter \citep{Freytag1996}. We set $f_{\rm CBM} = 0.02$ to be consistent with the equilibrium models used by \varghese. The value of the mixing-length parameter, $\alpha_{\rm MLT}$, is fixed to the solar-calibrated value of 1.713 by \cite{Choi2018}. The nuclear reactions included in the \texttt{pp\_cno\_extras\_o18\_ne22} network of \mesa are used. These reaction rates are from JINA REACLIB \citep{Cyburt2010} and NACRE \citep{Angulo1999}. \edit{The mass loss from stellar winds is computed from the prescription given by \citet[][scaling factor of 1]{Bjorklund2021}. } Finally, mixing by IGWs is activated once core-hydrogen burning is initiated. For now, we limit ourselves to non-rotating models, but we discuss rotational mixing later on in Sect.~\ref{sec:add_mix}. Our computational setup can be publicly accessed on Zenodo\footnote{\url{zenodo.org/records/14795326}}, as well as the \mesa models discussed in this work.

\subsection{Comparison to the hydrodynamical simulations}
\begin{figure*}
    \begin{subfigure}[b]{\textwidth}
        \centering
        \includegraphics[width=0.85\linewidth]{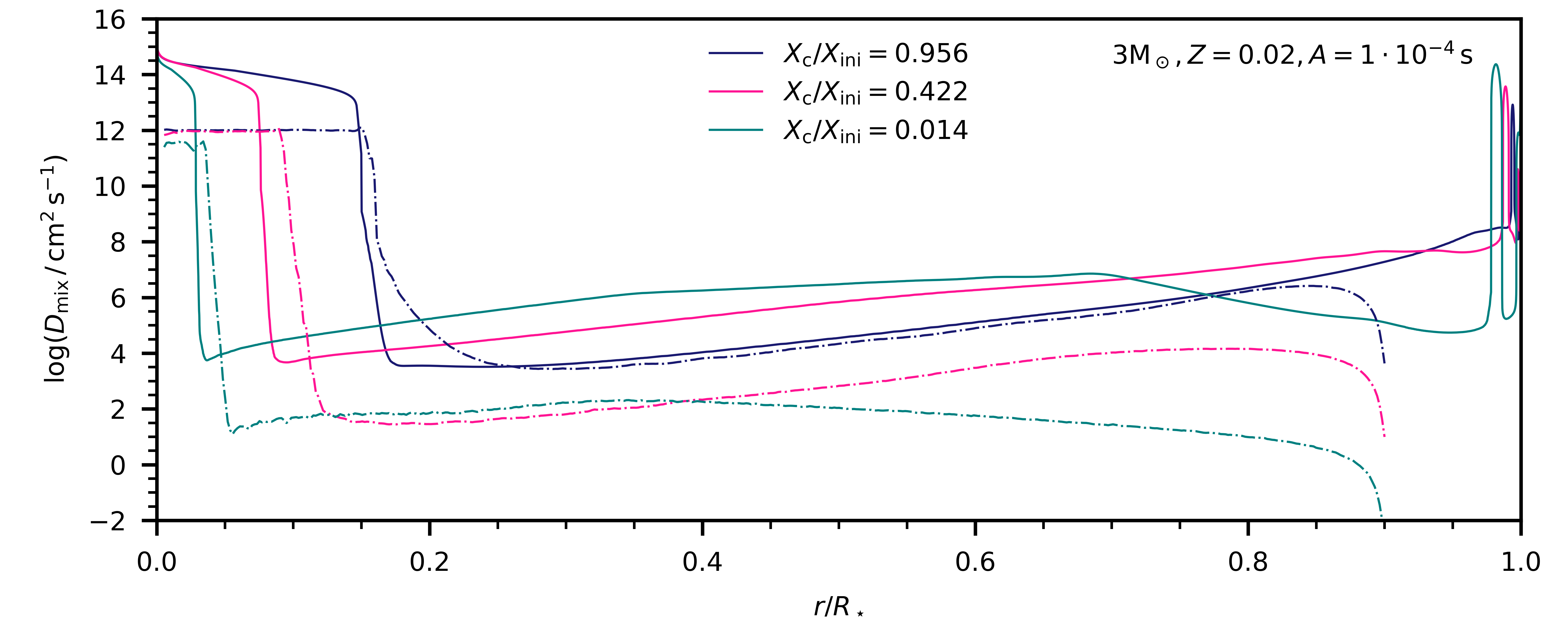}
    \end{subfigure}
    
    \begin{subfigure}[b]{\textwidth}
        \centering
        \includegraphics[width=0.85\linewidth]{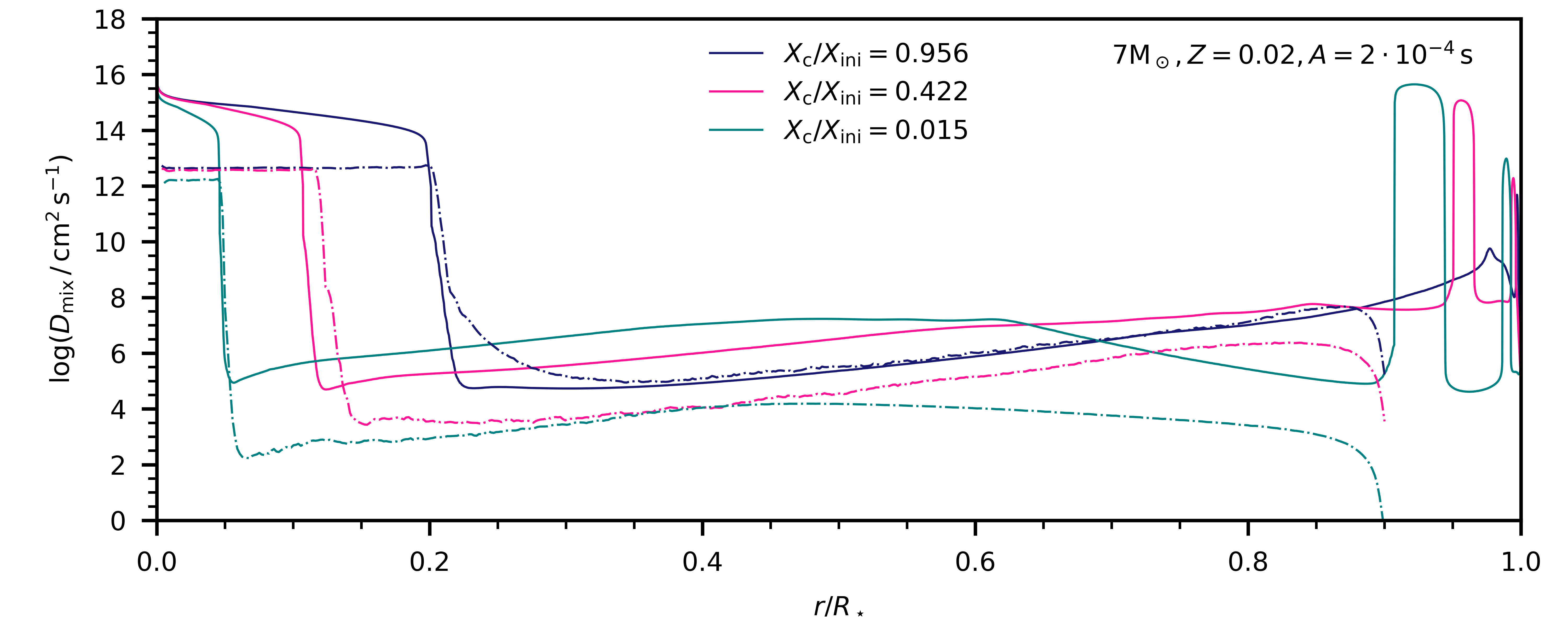}
    \end{subfigure}
    \caption{Predicted chemical diffusion coefficient from IGW mixing at different part of the main sequence (solid lines). Dashed-dotted lines indicate the results of the hydrodynamical simulations performed by \cite{Varghese2023}. The drop in the amplitude of the numerical profiles near the surface is due to the radial velocity being forced to zero at the top of the simulation domain (0.9R$_{\star}$ in these models) in the hydrodynamical simulations.  }
    \label{fig:D_IGW_profiles}
\end{figure*}

Figure \ref{fig:D_IGW_profiles} shows the total chemical diffusion coefficient including IGW mixing predicted from the approach discussed in Section~\ref{sec:num_approach} (solid lines), as well as the results from \varghese (dashed-dotted lines). As mentioned before, only the morphology of $D_{\rm IGW}$ (and not any offset) should be compared between predictions of the \mesa models and those of the hydrodynamical simulations. In general, the time evolution of the \edit{radial dependence of the local diffusion coefficient} is well described by our choice of the maximum wavenumber and the included wave frequencies. \edit{Based on the current state-of-the-art hydrodynamical simulations studying the chemical mixing by IGWs, we extrapolate our implementation up to 30\Msun, but we do note that the effects on the mixing by IGWs excited from possible surface convection zones are not characterized.}

\begin{figure*}
    \centering
    \includegraphics[width = \textwidth]{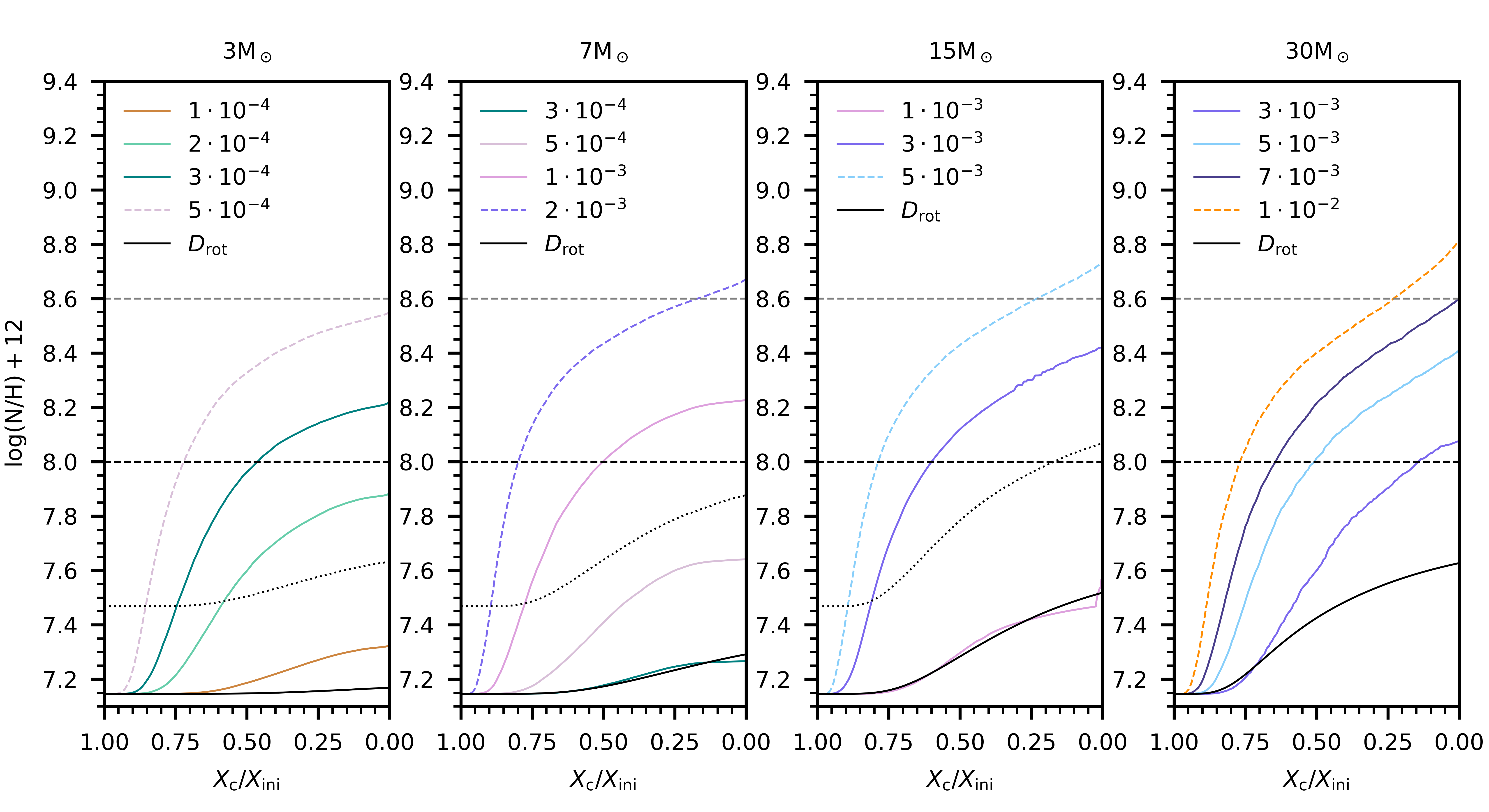}
    \caption{Predicted evolution of the N/H abundance ratio at the stellar surface for models with IGW mixing. Models are computed for a metallicity corresponding to the metallicity of the LMC. Different values of the legends indicate different values for the constant $A$ (in seconds). The black solid line indicate models with only rotational mixing included, starting at 25\percent of the initial \edit{Keplerian} critical rotation frequency. The black dotted lines show the SYCLIST models, also starting at 25\percent of the initial critical rotation frequency. The grey and black dashed horizontal lines indicate upper limits of stars in the LMC inferred by \cite{Martins2024} and \cite{Hunter2009}, respectively. The models plotted with a dashed line evolve towards higher effective temperatures during the main sequence. } 
    \label{fig:NH_LMC}
\end{figure*}

\begin{figure*}
    \centering
    \includegraphics[width = \textwidth]{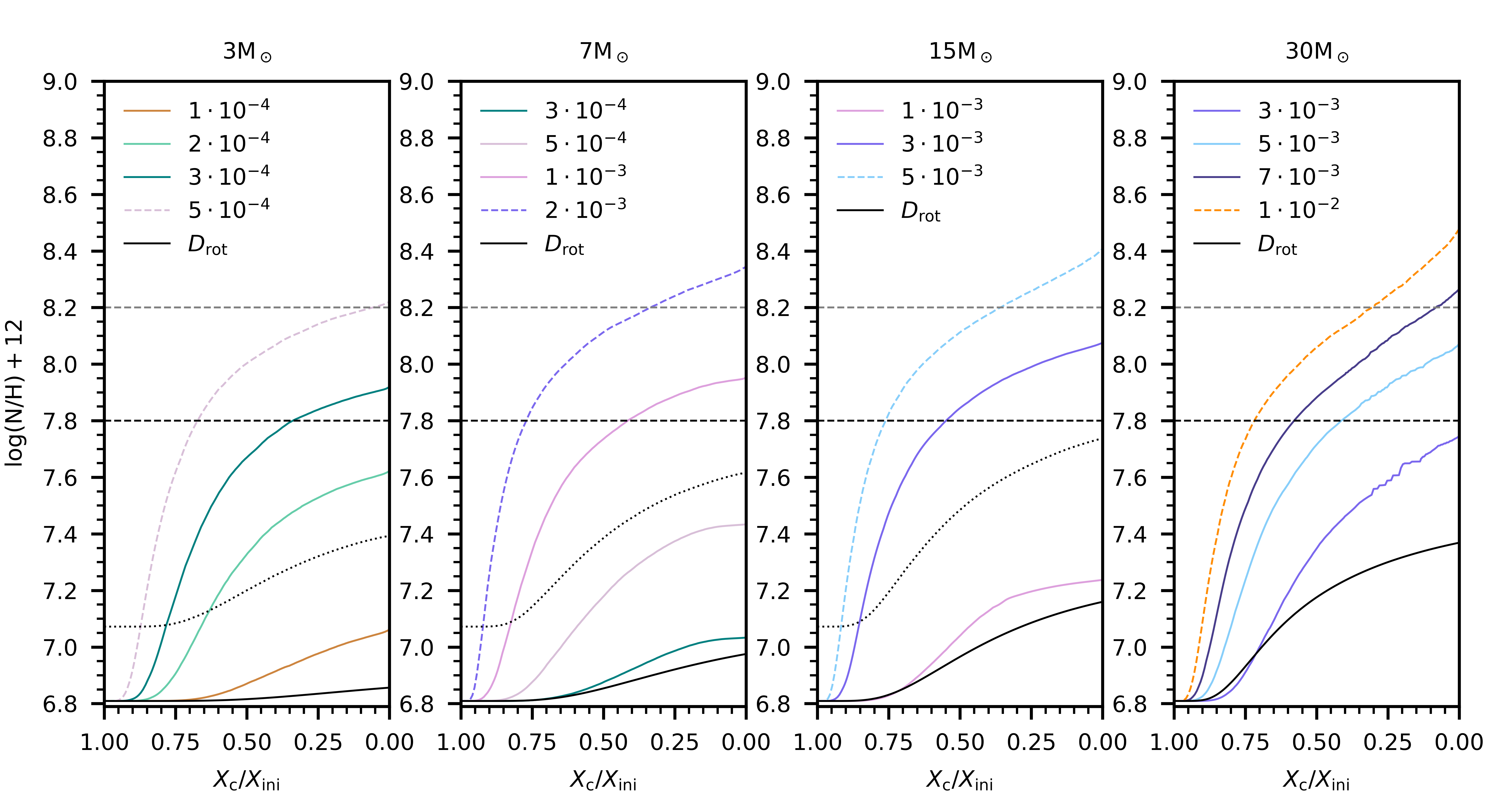}
    \caption{Same as Fig.~\ref{fig:NH_LMC}, but for a metallicity corresponding to the SMC. Observational limits are also for stars in the SMC.} 
    \label{fig:NH_SMC}
\end{figure*}

\section{Chemical evolution} \label{sec:results}
In the mass regime studied in this paper, the fusion of hydrogen into helium in the convective core is dominated by the carbon-nitrogen-oxygen (CNO) cycle. While these three elements function as catalysts, the $^{14}$N isotope builds up as a result of the different time scales of the nuclear reactions involved. Spectroscopic surveys of OB stars in the SMC and LMC have observed stars with significantly higher N/H fractions compared to the expected baseline, suggesting these stars have experienced efficient chemical mixing, transporting N from the core to the surface. In this section, we investigate what the value of the $A$ constant should be to reproduce the most nitrogen-enriched stars, assuming IGW mixing is the only form of mixing in the radiative envelope. 

Figures~\ref{fig:NH_LMC} and \ref{fig:NH_SMC} show the predicted evolution of the N/H ratio at the surface for different values of $A$, assuming $Z = 0.0064$ (LMC) and $Z = 0.0024$ (SMC), respectively (rotating models are discussed in Sect.~\ref{sec:add_mix}). We find IGW mixing rapidly increases the nitrogen surface abundance during the first half of the main sequence, but the efficiency diminishes during the second half, causing the N/H fraction to plateau off or increase less rapidly. In Appendix~\ref{ap:N-age}, versions of these figures are shown as a function of stellar age instead.

\begin{figure}
    \begin{subfigure}[b]{\columnwidth}
        \centering
        \includegraphics[width=\columnwidth]{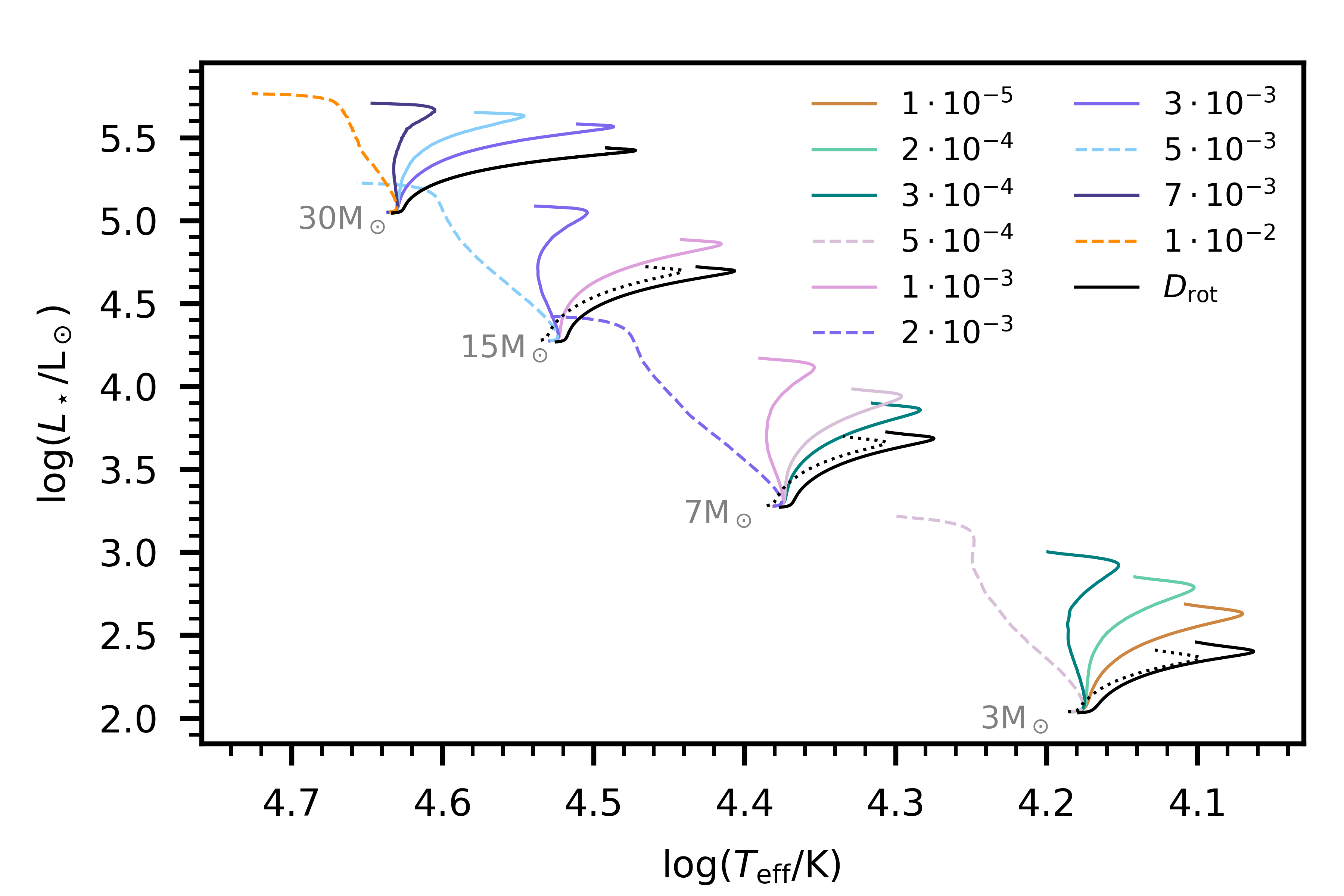}
        \caption{SMC metallicity ($Z = 0.0026$)}
    \end{subfigure}
    
    \begin{subfigure}[b]{\columnwidth}
        \centering
        \includegraphics[width=\columnwidth]{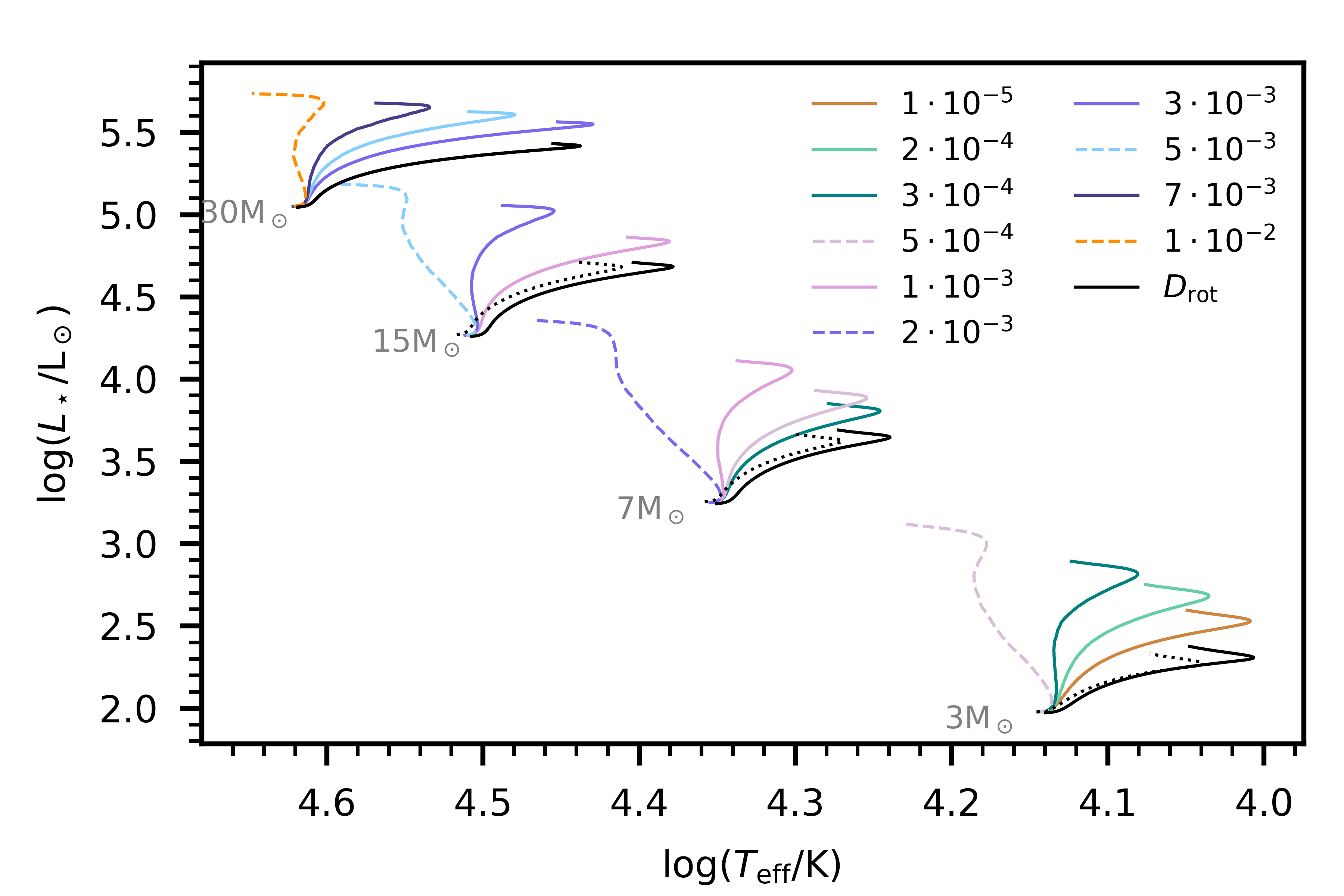}
        \caption{LMC metallicity ($Z = 0.0064$)}
    \end{subfigure}

    \begin{subfigure}[b]{\columnwidth}
        \centering
        \includegraphics[width=\columnwidth]{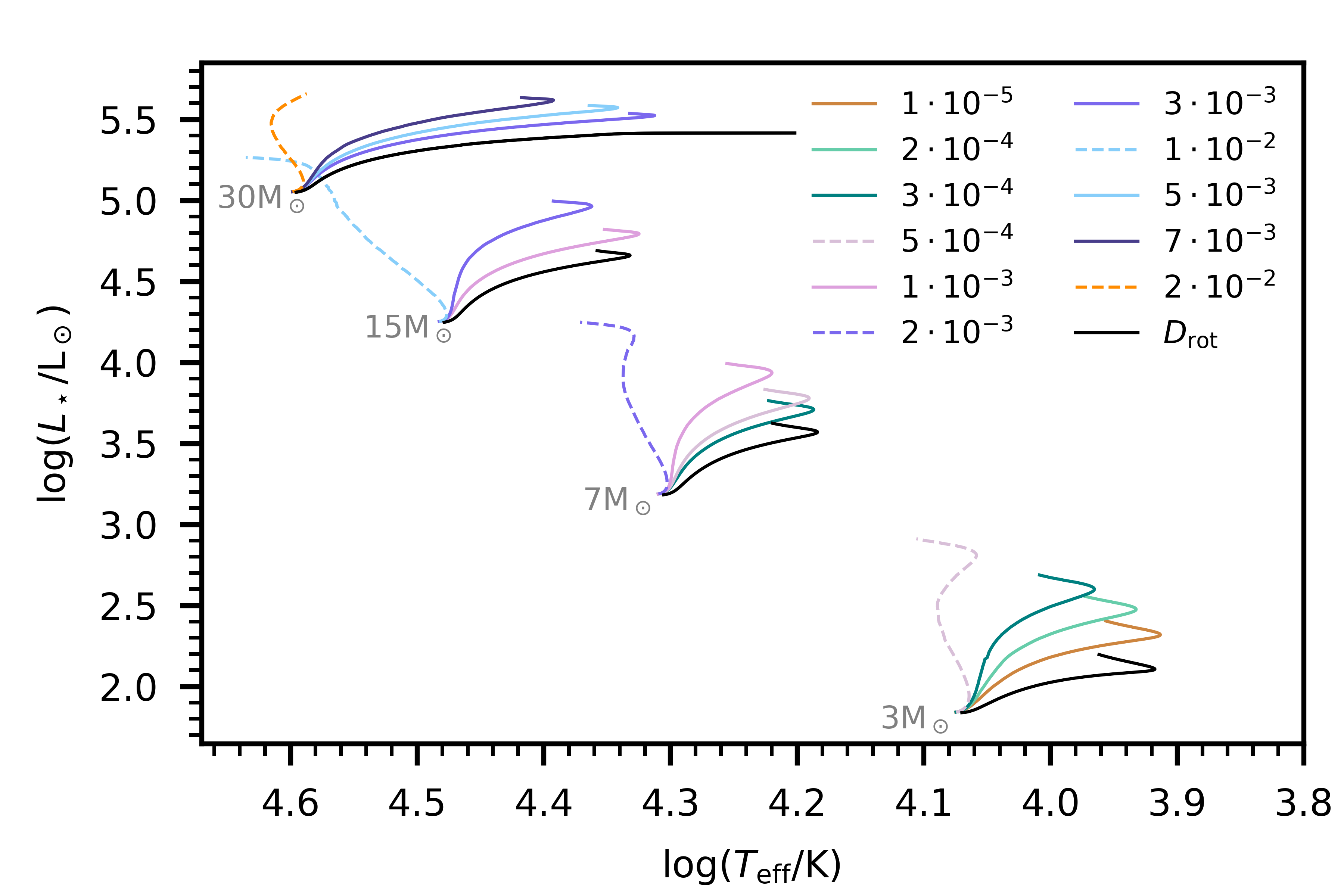}
        \caption{MW metallicity ($Z = 0.02$)}
    \end{subfigure}    
    \caption{Predicted evolution tracks for different values of the $A$ constant. Colour code and line style are the same as for Figs.~\ref{fig:NH_LMC} and \ref{fig:NH_SMC}.   }
    \label{fig:HRDs}
\end{figure}

Firstly, we note that there is an upper limit for $A$ above which the mixing is so efficient that the hydrogen mass fraction in the entire radiative envelope is significantly reduced by the nuclear burning in the core (i.e. quasi-chemically homogeneous evolution). This results in an increasing effective temperature during the main sequence evolution, as is shown by the dashed lines in Fig.~\ref{fig:HRDs}. The value for $A$ for which quasi-chemically homogeneous evolution occurs is around $5 \cdot 10^{-4}~{\rm s}$ for 3\Msun and increases with stellar mass to about $10^{-2}~{\rm s}$ for 30\Msun. For $Z = 0.02$, slightly higher values of $A$ are needed to observe quasi-chemically homogeneous evolution for 15 and 30\Msun, as shown in the bottom panel of Fig.~\ref{fig:HRDs}. The upper limits of the 1D models are orders of magnitude smaller than the value $\sim1~{\rm s}$ used in hydrodynamic simulations \citep{RogersMcElwaine2017, Varghese2023}. These upper limits hold for the entire metallicity range studied here. 

Secondly, we use the upper limits on N/H of single O-type stars from ULLYSES and XshootU surveys \citep{Vink2023,Martins2024} and of early-B stars from the VLT-FLAMES survey by \cite{Hunter2009} to acquire a \edit{calibrated} value for $A$. The survey by \cite{Martins2024} covers stars between roughly 20 and 40\Msun, and thus the 30\Msun model is the most representative here. Furthermore, we note that the baselines for the models presented in \cite{Hunter2009} are lower compared to ours. If we would instead of the upper limit on $\log({\rm N/H})$ itself use the upper limit on the difference with respect to the baseline, the black horizontal dashed line in Figs.~\ref{fig:NH_LMC} and \ref{fig:NH_SMC} would be roughly 0.3~dex higher in both cases. In order to roughly match the most nitrogen-enriched stars, we find the constant $A$ should scale with the stellar mass,
\begin{eqnarray} \label{eq:A}
    A/{\rm s} &=& -2.05 \cdot 10^{-7}(M_\star/{\rm M}_\odot)^3 + 1.14\cdot 10^{-5}(M_\star/{\rm M}_\odot)^2 \\ \nonumber
    &+& 7.75\cdot 10^{-5}(M_\star/{\rm M}_\odot) - 2.93 \cdot 10^{-5}.
\end{eqnarray}

\begin{figure*}
    \centering
    \includegraphics[width = \textwidth]{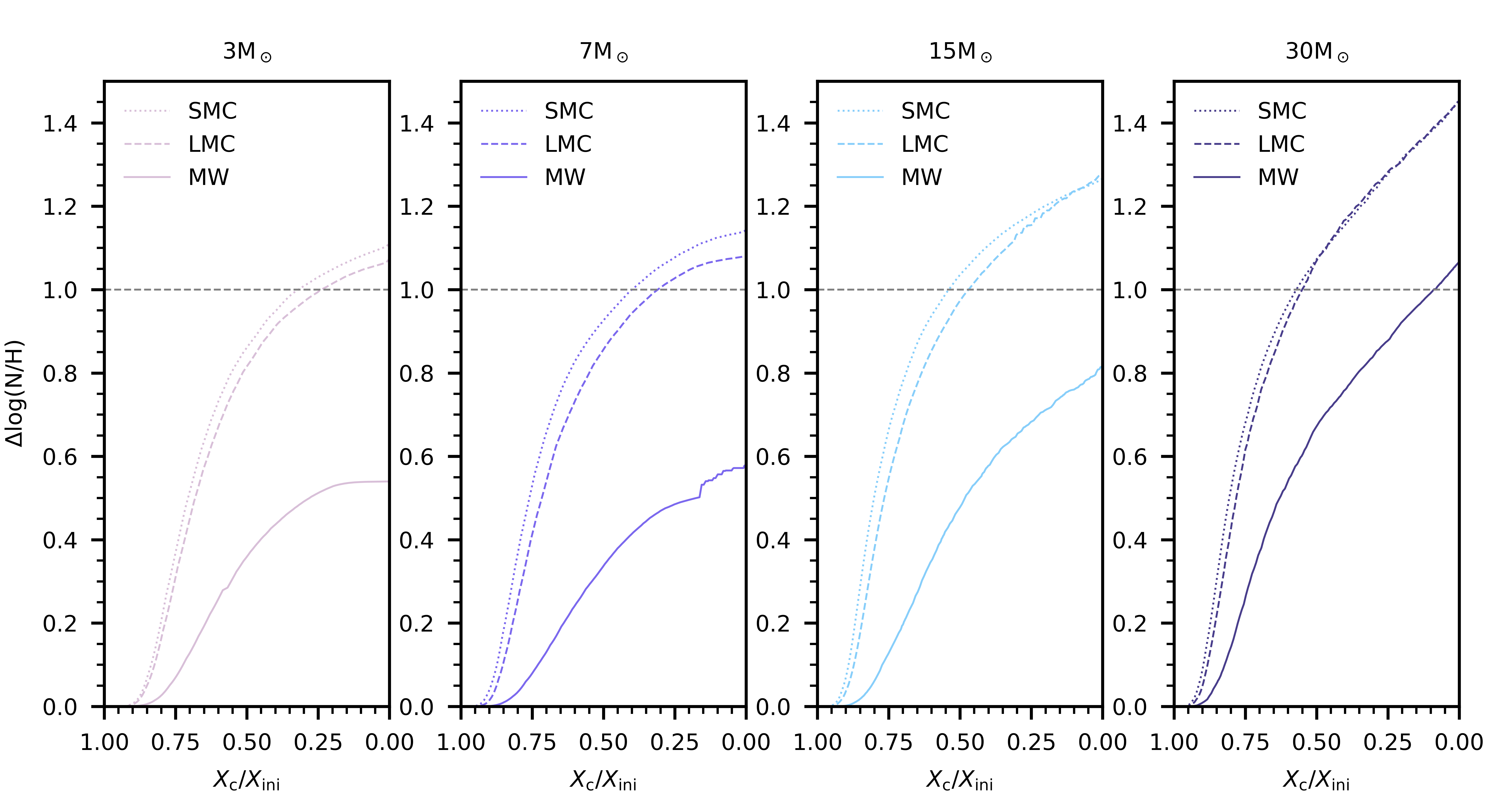}
    \caption{Predicted evolution of the N/H enrichment at the stellar surface with respect to the different baselines for models with IGW mixing. Colour code is the same as for Figs.~\ref{fig:NH_LMC} and \ref{fig:NH_SMC}. The grey horizontal dashed is the upper limit from \cite{Martins2024} for Galactic O-type stars. } 
    \label{fig:NH_metal}
\end{figure*}

Figure~\ref{fig:NH_metal} shows the predicted increase in N/H with respect to the baseline for SMC, LMC, and MW metallicity, where $A$ is set according to Eq.~(\ref{eq:A}). We find similar values for the models with SMC and LMC metallicity, and weaker nitrogen enrichment for the MW models. The survey by \cite{Hunter2009} finds $\Delta \log({\rm N/H}) \lesssim$ 0.2~dex for Galactic early-B stars, which is less than what is predicted from the models of 7 and 15\Msun. On the other hand, the survey by \cite{Martins2024} does find $\Delta \log({\rm N/H}) \lesssim$ 1~dex, which is in line with the value the 30\Msun model reaches towards the end of the main sequence (solid line in right-most panel of Fig.~\ref{fig:NH_metal}). 

\begin{figure*}
    \centering
    \includegraphics[width=0.85\textwidth]{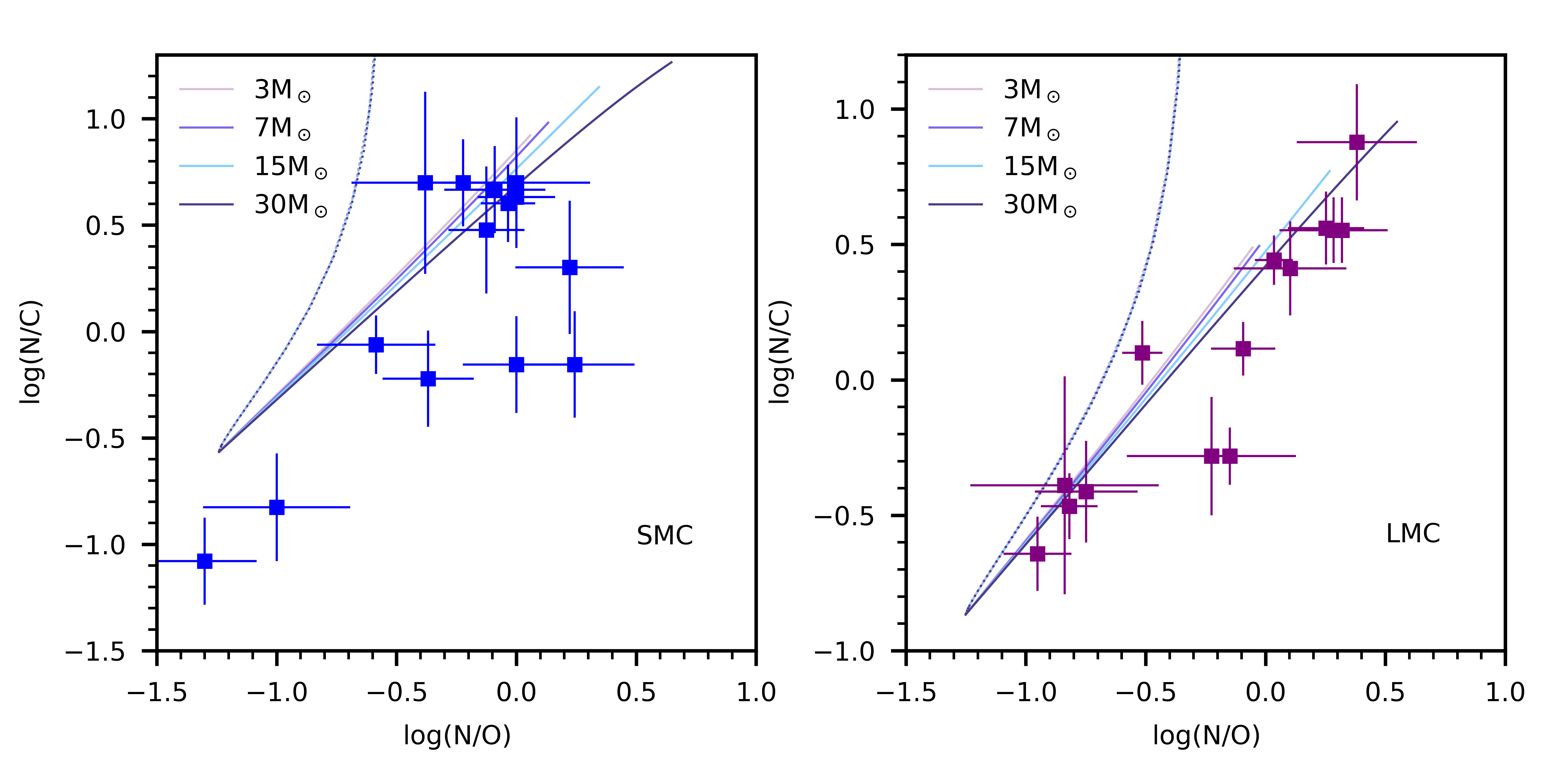}
    
    \caption{Predicted coverage of the ratios N/O and N/C at the surface (solid lines) for models with IGW mixing. In this diagram, stars evolve from the lower left corner to the upper right corner. Data points indicate the measurements from \cite{Martins2024}. The dotted lines indicate the ratios in the convective core.   }
    \label{fig:NO-NC}
\end{figure*}

In addition to the N/H ratio, we show in Fig.~\ref{fig:NO-NC} the predicted ratios of N/O versus N/C for models with IGW mixing from ZAMS to TAMS, where $A$ is again set following Eq.~(\ref{eq:A}). Increasing $A$ results in more extreme abundance values between ZAMS and TAMS. For comparison, individual abundance measurements by \cite{Martins2024} are shown for O-type stars in the SMC and LMC. We find that our choice for $A$ does cover all measured ratios for the most massive models. In case of the SMC, some stars have an offset in the N/C ratio compared to the models, pointing to a slightly different baseline.

\section{Convective core mass} \label{sec:core_masses}
\begin{figure}
    \centering
    \includegraphics[width=\columnwidth]{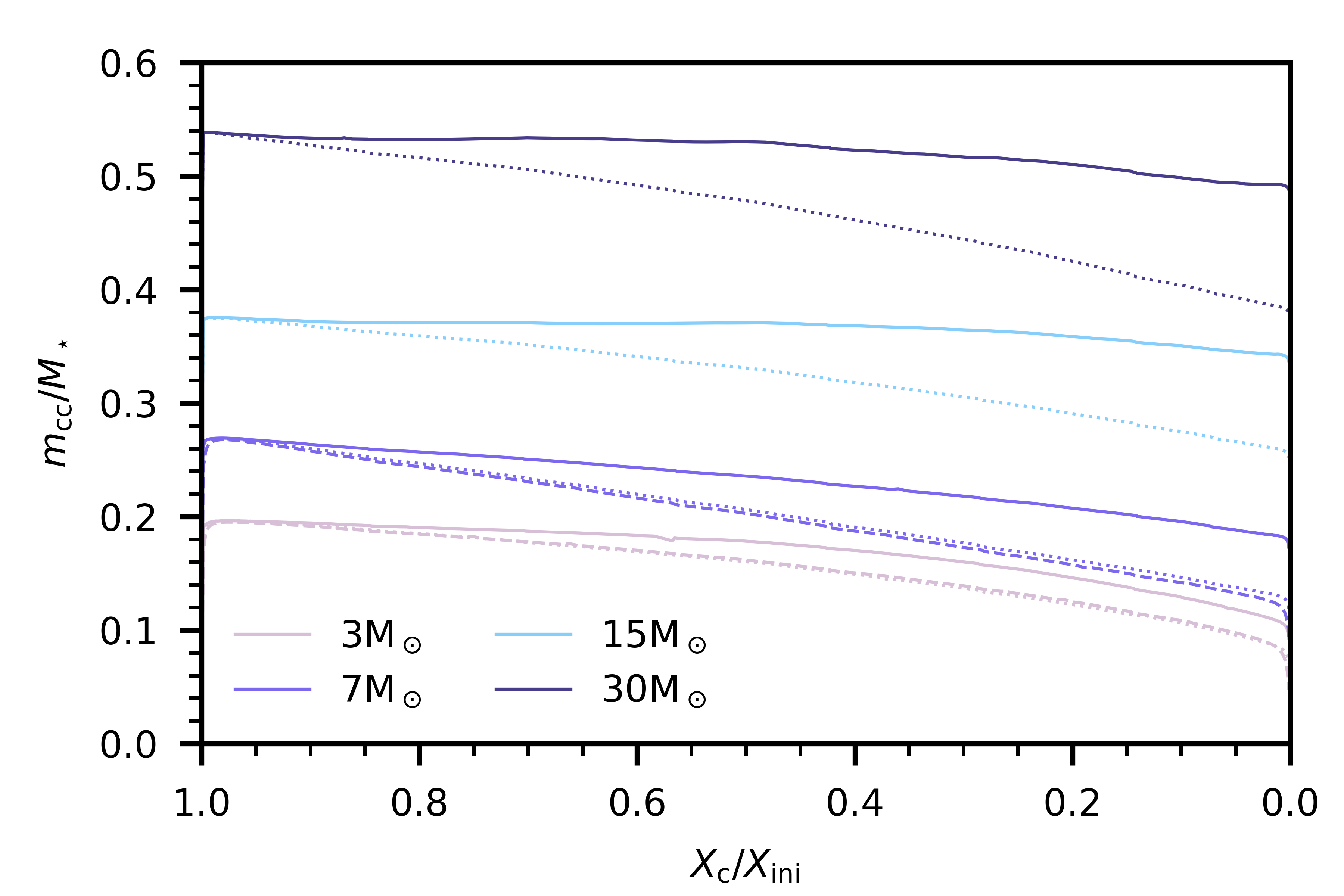}
    \caption{Predicted evolution of the convective core mass for models with IGW mixing and $Z = 0.02$. The solid lines correspond to models where the value of $A$ has been set according to Eq.~(\ref{eq:A}). The dotted lines indicate the lowest values per mass considered in Fig.~\ref{fig:NH_LMC}. The two dashed lines indicate models with $f_{\rm CBM} = 0.005$ instead of 0.02, where $A = 2 \cdot 10^{-4}{\, \rm s}$ for 3\Msun and $A = 5 \cdot 10^{-4}{\, \rm s}$ for 7\Msun. In case of the 3\Msun model, the dotted and dashed lines coincide on the plot.   }
    \label{fig:mcc}
\end{figure}

The chemical mixing induced by IGWs can lead to more massive convective cores as efficient mixing introduces additional hydrogen to the core and thereby prolonging a star's main-sequence life time. Asteroseismic measurements of convective core masses in massive stars have shown that the observations require more massive cores than what is predicted from models \citep{Johnston2021}. In Fig.~\ref{fig:mcc}, we show the predicted evolution of the fractional convective core mass along the main sequence for models with IGW mixing. The solid lines correspond to the models with a value of $A$ according to Eq.~(\ref{eq:A}). The dotted lines indicate models with the lowest values of $A$ presented in this work, that is, minimal mixing, which results in no observable nitrogen enrichment. The models that reproduce the most nitrogen-enriched stars also predict these stars to have significantly more massive cores, up to $\sim$40\percent larger at the TAMS compared to models with inefficient mixing. 

Typically, the core masses in stellar evolution models are enlarged by including CBM, which is in the models presented here controlled by the parameter \fov. In Fig.~\ref{fig:mcc}, we also show the predicted core mass for models with $f_{\rm CMB} = 0.005$ (3 and 7\Msun), instead of 0.02, but have a higher value of $A$ (dashed lines). These models for $f_{\rm CMB} = 0.005$ (0.02) indicated by the dashed (dotted) lines have $A = 2 \cdot 10^{-4}{\, \rm s}$ ($1 \cdot 10^{-4}{\, \rm s}$) and $A = 5 \cdot 10^{-4}{\, \rm s}$ ($2 \cdot 10^{-4}{\, \rm s}$) for 3 and 7\Msun, respectively.   It can be seen that a higher value of $A$ can compensate for a lower value of \fov in terms of obtaining the same core mass. The value of \fov, however, has no significant influence on the surface abundances. While asteroseismically modelled slowly pulsating B-type stars, ranging between 3 and 9\Msun, seem to have a decreasing fractional core mass as they evolve \citep{Pedersen2021}, these stars also do not show any significant nitrogen-enrichment \citep{Gebruers2021}. Similar behaviour of the core mass is observed in $\beta$~Cephei stars \citep{Fritzewski2024} that span the 10 to 20\Msun mass range. Future asteroseismic measurements of core masses of stars that are nitrogen-enriched are needed to confirm whether these stars indeed have a more massive core. Such modelling efforts should then also rely on models that include IGW mixing.   

\section{Additional sources of mixing} \label{sec:add_mix}
 \edit{A significant fraction of the sample of stars shows nitrogen-enrichment at the surface have rotation velocities that are too small to explain the observed abundances by rotational mixing alone \citep{Hunter2009, Brott2011b}. } In this section, we quantify the relative strength of IGW mixing compared to other commonly used mechanisms and implementations for chemical mixing in the radiative envelope.

\subsection{Rotational mixing following Heger et al.}
The first widely adapted diffusive scheme for rotational mixing (and implemented in \mesa) is based on the total contribution of six (magneto)hydrodynamical processes; dynamical shear instability, secular shear instability, Eddington-Sweet circulation, Goldreich-Schubert-Fricke instability, Solberg-H\o iland instability, and Spruit-Tayler dynamo. The total sum of these diffusion coefficients should be lower for the chemical mixing, $D_{\rm rot}$, compared to the diffusion coefficient (viscosity), $\nu$, for angular momentum transport to prevent the star from becoming fully mixed. Therefore, the total chemical diffusion coefficient is scaled with factor $f_{\rm c} = D_{\rm rot}/\nu$ \citep{Pinsonneault1989}. Moreover, for the latter two processes, it is debated that they do not contribute to the chemical diffusion \citep[cf.][]{Brott2011a}, \edit{and therefore we set $f_{\rm c} =0$ for these two processes}. The second free parameter in this mixing scheme is the sensitivity of rotational mixing to the chemical gradient, which is parameterised by multiplying the chemical gradient by a factor $f_\mu$ in the expressions used to determine the chemical diffusion coefficients. There exist a degeneracy between $f_{\rm c}$ and $f_\mu$, and thus these parameters should be calibrated as a set. \cite{Heger2000} observationally calibrated the value of $f_\mu = 0.05$ for $f_{\rm c} = 1/30$, and these values are commonly used in the literature. Here, we also use these values, but note that these parameters are not well constrained. The models are initialised with a uniform rotation frequency of 25\percent of the critical (Keplerian) rotation frequency at the ZAMS. \edit{We choose this rotation rate to represent a moderate rotator for which the centrifugal deformation is still small.} These models are shown in black in Figs.~\ref{fig:NH_LMC} and \ref{fig:NH_SMC}, for which it can been seen that with this implementation of rotational mixing \edit{(and no IGW mixing)}, the most nitrogen-enriched stars cannot be explained. To reach the upper limits for 15 and 30\Msun, an initial rotation of ~50\percent critical is needed, but then these models reach critical rotation during the main sequence. Therefore, (this implementation of) rotational mixing, \edit{using the aforementioned values of the free parameters}, is not viable as the only source of mixing within the mass regime studied here.
 
\subsection{Rotational mixing following Zahn}
The second scheme of rotational mixing that is widely used is based on the works of \cite{Zahn1992} and \cite{ChaboyerZahn1992}. \edit{In this framework, we adopt the coefficient related to vertical shear mixing given by,}
\begin{equation}
    D_{\rm shear} = K\left(\frac{r}{N} \frac{{\rm d}\Omega}{{\rm d}r} \right)^2,
\end{equation}
where $K$ is the thermal diffusivity, $N$ the Brunt-V\"ais\"al\"a frequency, and $\Omega$ the local angular velocity. In addition to $D_{\rm shear}$, a diffusion coefficient attributed to the meridional circulation and horizontal turbulence, commonly named $D_{\rm eff}$, is added. However, here we do not include $D_{\rm eff}$, as \cite{Mombarg2023-ester} argue based on self-consistent 2D rotating stellar evolution models that chemical transport due to meridional circulation is extremely slow. Furthermore, for the transport of angular momentum, it has been shown that using a viscosity as predicted by \cite{Zahn1992} does not explain asteroseismic measurements of pulsating F-type stars \citep{Ouazzani2019}, and that magnetic processes should also be taken into account \citep{Moyano2023}. Here, we simplify by assuming a constant viscosity (in the envelope and throughout the evolution) for the transport of angular momentum. We use the lower limit derived by \cite{Mombarg2023-am} for F-type stars, namely $\nu = 10^7\,{\rm cm}^2\,{\rm s}^{-1}$. We observe no nitrogen enrichment in any of the models when assuming the same fraction of critical rotation at the ZAMS. Figures~\ref{fig:NH_LMC} and \ref{fig:NH_SMC} show the predictions from (interpolated) SYCLIST models \citep{Georgy2013}, for 3, 7, and 15\Msun, starting with a rotation equal to 25\percent of the critical rotation frequency\footnote{\edit{The SYCLIST models use the Roche definition of the critical rotation frequency, $\Omega/\Omega_{\rm crit, Kepler} = 0.25$ corresponds to $\Omega/\Omega_{\rm crit, Roche} = 0.46$. These models have been calibrated to average nitrogen abundances, assuming an initial rotation rate $\Omega/\Omega_{\rm crit, Roche} = 0.568$ \citep{Ekstrom2012}.}}. The models have the same metallicities as the \mesa models presented in this paper, but a different baseline. We note that for this initial rotation velocity the evolution of the surface nitrogen enrichment predicted by these models is \edit{slightly higher compared} to what is predicted by the \mesa models in the previous section.

\edit{We conclude that IGW mixing, assuming $A$ scales according to Eq.~(\ref{eq:A}), can reproduce stars that show signs of efficient mixing, while their rotation frequency is too low for rotational mixing to be sufficiently efficient \citep[e.g.][]{Brott2011b}.}
Gaining an understanding on the different factors contributing to the value of the $A$ constant can provide a more definite answer to the question whether the mixing induced by IGWs dominates over mixing induced by rotation. 

\subsection{Microscopic diffusion}
\begin{figure}
    \centering
    \includegraphics[width = \columnwidth]{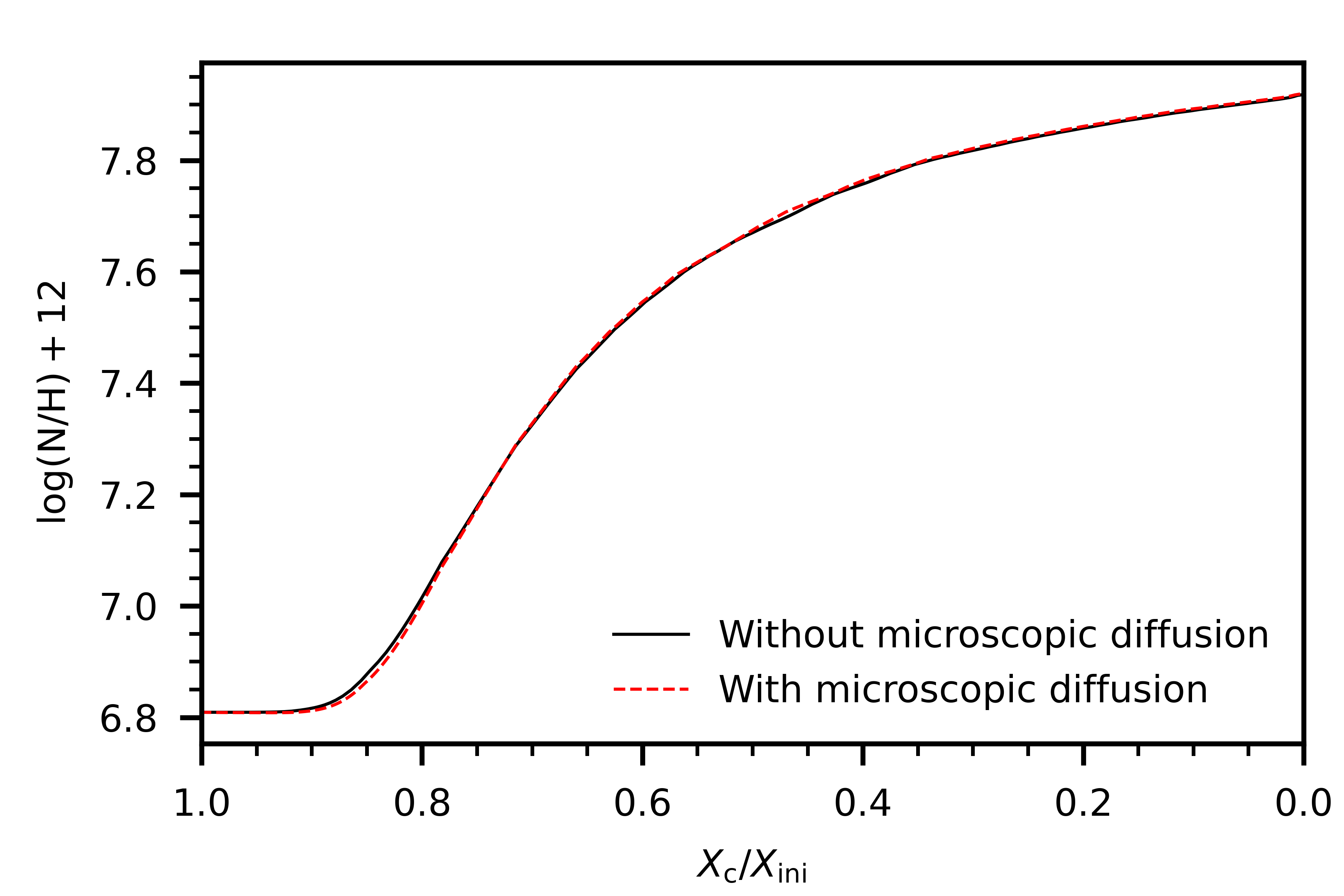}
    \caption{Predicted evolution of the N/H ratio at the stellar surface with and without microscopic diffusion included. The models have a mass of 3\Msun, a metallicity of $Z = 0.0026$, and a value of $A = 3 \cdot 10^{-4}\,{\rm s}$.}
    \label{fig:NH_radlev}
\end{figure}

Besides chemical mixing on a macroscopic level, mixing also occurs on a microscopic level \citep{Michaud2015}. We computed also a model with microscopic diffusion (i.e. gravitational settling, concentration diffusion, thermal diffusion, and radiative levitation) included. As shown in Fig.~\ref{fig:NH_radlev} for an example for the 3\Msun model, SMC metallicity, the resulting evolution of the surface nitrogen abundance is not significantly impacted by microscopic diffusion when IGW mixing is active. We note that while we neglect microscopic diffusion for the evolution of the CNO elements at the surface, this does not imply this type of mixing is irrelevant is the mass regime we study here. For example, heavier elements like iron and nickel experience a significantly stronger radiative acceleration and the accumulation of these elements around the iron-bump at 200\,000~K could alter the excitation of eigenmodes in slowly-pulsating B-type stars, even with IGW mixing present \citep{Rehm2024}. 

\section{Conclusions} \label{sec:conclusions}
In this study, we have implemented a scheme for IGW mixing that is based on 2D hydrodynamical simulations by \cite{Varghese2023} into the 1D stellar structure and evolution code \mesa. Using this novel implementation, we have studied the evolution of the N/H surface abundance ratio for models of 3, 7, 15, and 30\Msun, and for metallicities corresponding to the SMC, LMC, and Milky Way. We investigated different values of the constant $A$ that relates the squared IGW velocity to the chemical diffusion coefficient (cf. Eq.~(\ref{eqn:param_D})), and we provided estimates of its value such that the highest measured N/H fractions in the surveys of \cite{Hunter2009} and \cite{Martins2024} can be reproduced. This requires $A$ to scale with stellar mass, and we provide a relation in Eq.~(\ref{eq:A}). This study therefore provides observational limits on the value of $A$ that serve as a guideline for future theoretical studies on the efficiency of IGW mixing. The values of $A$ that we derived here correspond to a maximised effect from a range of contributing frequencies based on \cite{Varghese2023}. A possible future improvement would be to only account for the dominant frequency, once these dominant frequencies can be reliably predicted at any given mass and any given age. 

Furthermore, we have compared the mixing efficiency of our novel IGW mixing implementation with commonly used implementations of rotational mixing and conclude that mixing by IGWs can indeed be the prevalent form of mixing in the radiative envelope of massive stars if the star starts the main sequence as a slow or moderate rotator. Indeed, some stars appear to have high N/H fractions, yet a low surface rotation velocity. We do recognise, however, that the efficiency of rotational mixing over time is strongly dependent on the efficiency of angular momentum transport, which is also ill-understood (see e.g. \citealt{Aerts2019-ARAA}). Moreover, IGWs can induce zones of strong shear that can then experience additional mixing driven by shear instabilities. Recently, \citet{Varghese2024} studied the influence of rotation on wave mixing considering a 7\Msun star rotating up to 35 and 69\percent of the critical rotation velocity at ZAMS and MAMS, respectively. They found the wave mixing to decrease with rotation due to the influence of rotation on convection which subsequently affects the amplitude with which waves are generated at the convective-radiative interface and its propagation in the radiation zone, particularly in older stars. Their studies suggest that accounting for the effect of rotation on convection could provide a better constraint on the mixing by IGWs in the radiative envelope.

\begin{acknowledgements}
  The authors thank Conny Aerts, Tamara Rogers, Philipp Edelmann, Dominic Bowman, Riccardo Vanon and St\'ephane Mathis for the useful discussions on the work. \edit{The authors also thank the anonymous referee for the comments that have improved the clarity of the manuscript.}
  The research leading to these results has received funding from the French Agence Nationale de la Recherche (ANR), under grant MASSIF (ANR-21-CE31-0018-02) and from the European Research Council (ERC) under the Horizon Europe programme (Synergy Grant agreement N$^\circ$101071505: 4D-STAR).  While partially funded by the European Union, views and opinions expressed are however those of the authors only and do not necessarily reflect those of the European Union or the European Research Council. Neither the European Union nor the granting authority can be held responsible for them. RPR was supported by the STFC grant ST/W001020/1. The hydrodynamical simulations were carried out on the DiRAC Data Intensive service at Leicester (DIaL), operated by the University of Leicester IT Services, which forms part of the STFC DiRAC HPC Facility (\url{www.dirac.ac.uk}), funded by BEIS capital funding via STFC capital grants ST/K000373/1 and ST/R002363/1 and STFC DiRAC Operations grant ST/R001014/1. The computational resources and services used in
this work were provided by the VSC (Flemish Supercomputer Center), funded by
the Research Foundation – Flanders (FWO) and the Flemish Government department EWI. This research made use of the \texttt{numpy} \citep{Harris2020} and \texttt{matplotlib} \citep{Hunter2007} \texttt{Python} software packages.   
\end{acknowledgements}

%
%
\bibliographystyle{aa} 
\bibliography{main} 

\begin{appendix}
\section{Initial wave velocity} \label{ap:uv0}
\edit{
The fraction of the convective velocity, $v_{\rm MLT}$, that is transmitted to the vertical velocity of the IGW is expressed by \cite{Press1981} as,
\begin{equation}
    u_{\rm v, 0} = \frac{F k_{\rm h}}{\rho \sqrt{N^2 - \omega^2}}.
\end{equation}
The convective flux, $F$, following \cite{Lecoanet2013} is calculated as,
\begin{equation}
    F = \rho v_{\rm MLT}^3 \frac{\lambda}{h_{P}},
\end{equation}
where $h_P$ is the local pressure scale height, and $\lambda$ indicates the radial length scale of the wave \citep[][their Eq.~(142)]{Ahuir2021},
\begin{equation}
    \lambda = \omega^{2/3}(\ell(\ell+1))^{-1/3} \left| \frac{{\rm d} N^2}{{\rm d}r}\right |_{r_{\rm int}}^{-1/3} r_{\rm int}^{2/3}.
\end{equation}
Here, $r_{\rm int}$ is the radial coordinate of the convective interface. Numerically speaking, defining this $r_{\rm int}$ is difficult and the resulting $u_{\rm v, 0}$ is very sensitive to the definition as $N^2$ varies rapidly near the convective interface. If we define $r_{\rm int}$ as the radius where the Brunt-V\"ais\"al\"a frequency becomes larger than the largest wave frequency we take into account ($N > 14 \mu{\rm Hz}$) the lowest fraction for $u_{\rm v, 0}/v_{\rm MLT}$ we obtain between 4 and 14$\mu{\rm Hz}$ is $\sim 0.02$.
}
\section{Surface nitrogen abundance versus age} \label{ap:N-age}
In this appendix, we show versions of Figs.~\ref{fig:NH_LMC} and \ref{fig:NH_SMC} with the stellar age on the abscissa instead of $X_{\rm c}/ X_{\rm ini}$. 

\begin{figure*}
    \centering
    \includegraphics[width = 0.87\textwidth]{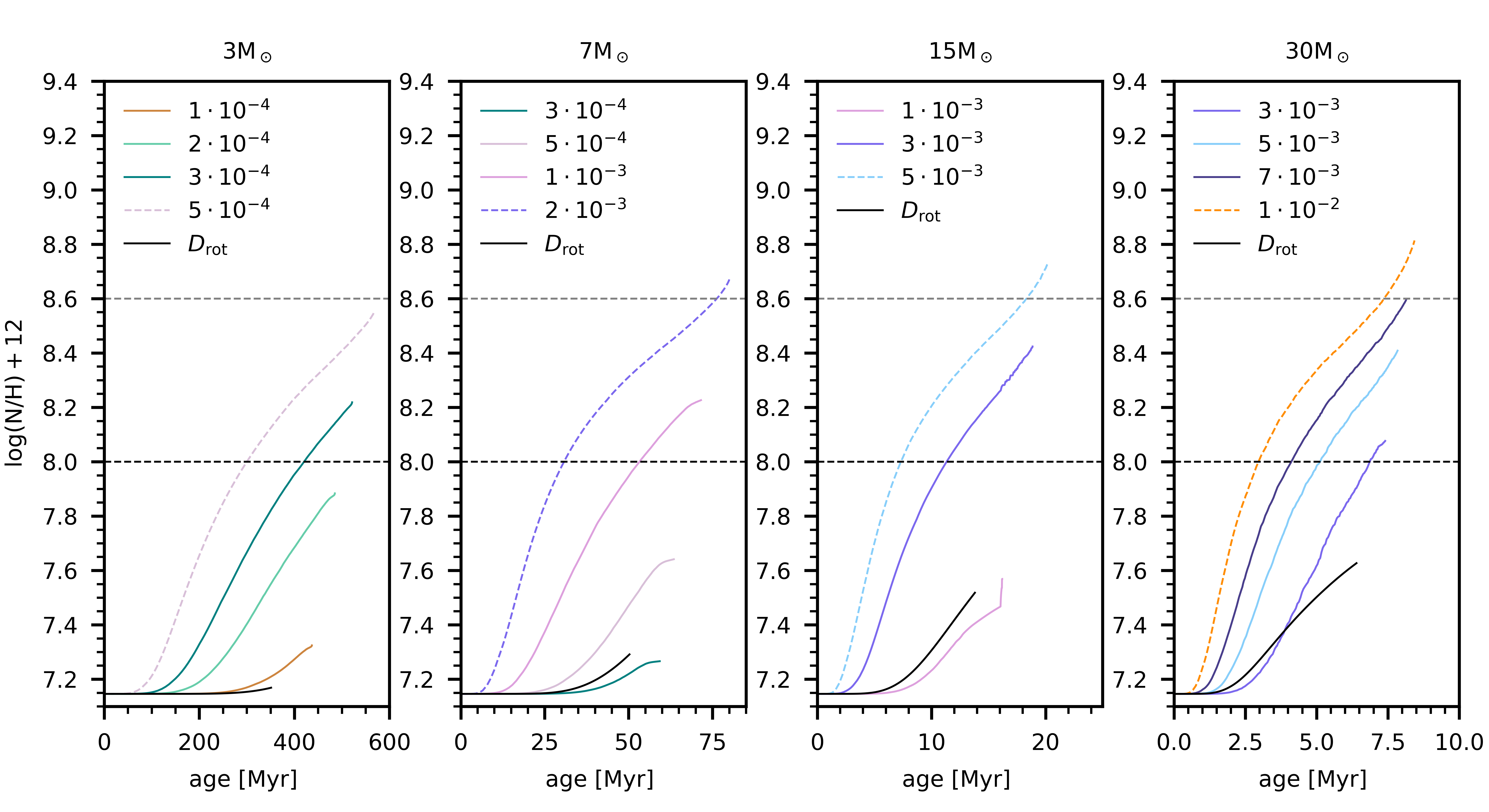}
    \caption{Predicted evolution of the N/H abundance ratio at the stellar surface for models with IGW mixing. Models are computed for a metallicity corresponding to the metallicity of the LMC. Different values of the legends indicate different values for the constant $A$ (in seconds). The black solid line indicate models with only rotational mixing included, starting at 25\percent of the initial critical rotation frequency. The black dotted lines show the SYCLIST models, also starting at 25\percent of the initial critical rotation frequency. The grey and black dashed horizontal lines indicate upper limits of stars in the LMC inferred by \cite{Martins2024} and \cite{Hunter2009}, respectively. The models plotted with a dashed line evolve towards higher effective temperatures during the main sequence. } 
    \label{fig:NH_LMC_age}
\end{figure*}

\begin{figure*}
    \centering
    \includegraphics[width = 0.87\textwidth]{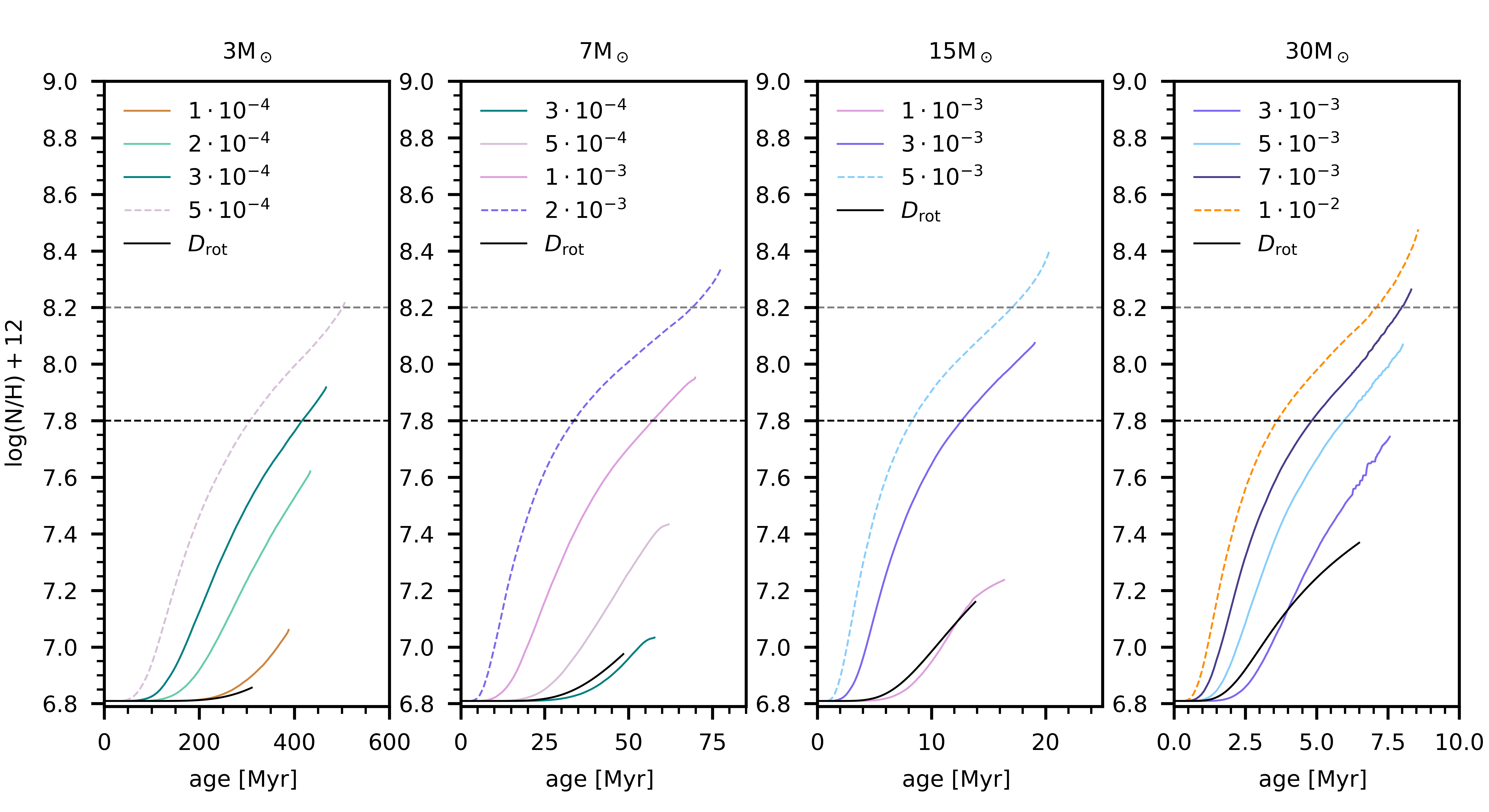}
    \caption{Same as Fig.~\ref{fig:NH_LMC_age}, but for a metallicity corresponding to the SMC. Observational limits are also for stars in the SMC.} 
    \label{fig:NH_SMC_age}
\end{figure*}

\end{appendix}

\end{document}